\documentclass[final,9pt]{elsarticle}

\usepackage[shortlabels]{enumitem}
\usepackage{amssymb}
\usepackage{amsmath}
\usepackage{amsthm}
\newtheorem{theorem}{Theorem}
\newtheorem{lemma}{Lemma}
\newtheorem{definition}{Definition}

\journal{arXiv}
\begin{document}

\begin{frontmatter}



\title{Turing instabilities for three interacting species}


\author[inst1]{Vit Piskovsky}
%\ead{vit.piskovsky@maths.ox.ac.uk} % email address
\affiliation[inst1]{organization={Mathematical Institute, University of Oxford},%Department and Organization
            addressline={Woodstock Road},
            city={Oxford},
            postcode={OX2 6GG},
            country={UK, vit.piskovsky@maths.ox.ac.uk}}

\begin{abstract}
In this paper, I prove necessary and sufficient conditions for the existence of Turing instabilities in a general system with three interacting species. Turing instabilities describe situations when a stable steady state of a reaction system (ordinary differential equation) becomes an unstable homogeneous steady state of the corresponding reaction-diffusion system (partial differential equation). Similarly to a well-known inequality condition for Turing instabilities in a system with two species, I find a set of inequality conditions for a system with three species. Furthermore, I distinguish conditions for the Turing instability when spatial perturbations grow steadily and the Turing-Hopf instability when spatial perturbations grow and oscillate in time simultaneously.
\end{abstract}


%\item It is possible to determine when a general reaction-diffusion system with three interacting species forms Turing patterns.
%\item Turing patterns form due to the instability of the homogeneous stationary state to spatial perturbations.
%\item Two instability types can be distinguished: Turing instability suggests the formation of stationary patterns, while Turing-Hopf instability suggests the formation of dynamic patterns.


\end{frontmatter}

%\linenumbers

%% Introduction
\section{Introduction}\label{sec:intro}
Assume that $\mathbf{n}^*$ is a unique stable steady state of the reaction system
\begin{equation}\label{eq:de_temporal}
    \frac{\mathrm{d}n_i(t)}{\mathrm{d}t} = f_i(\mathbf{n}(t))
\end{equation}
with species $i=1, \dots, N$ of concentrations $\mathbf{n}=(n_1,\dots,n_N)$. The linear stability of $\mathbf{n}^*$ as a steady state for the reaction-diffusion system
\begin{equation}\label{eq:pde_fix_diff}
    \partial_t n_i(t,\mathbf{x}) = d_i\partial_\mathbf{x}^2 n_i(t,\mathbf{x})+f_i(\mathbf{n}(t,\mathbf{x}))
\end{equation}
with species $i$ diffusing at rate $d_i>0$ in the space $\mathbf{x} \in \mathbb{R}^N$ is determined by the eigenvalues of the matrix
\begin{equation}\label{eq:jacobian}
    J(k^2)=J-k^2\mathrm{diag}(d_1, \dots,d_N),
\end{equation}
where the matrix $J=\partial \mathbf{f}/\partial \mathbf{n}|_{\mathbf{n}^*}$ is the Jacobian for the reaction system \eqref{eq:de_temporal} and, by assumption, has eigenvalues with negative real parts. We make the following definitions to analyze whether perturbation modes of wavelength $k$ can grow and oscillate.
\begin{definition}
    The spatially uniform steady-state $\mathbf{n}^*$ of \eqref{eq:pde_fix_diff} admits an instability if there is a wavenumber $k$ such that $J(k^2)$ has an eigenvalue with a positive real part. Furthermore, if any eigenvalue of $J(k^2)$ with a positive real part is real for any wavenumber $k$, we call the instability a Turing instability. Otherwise, if some eigenvalues of $J(k^2)$ with a positive real part are not real at some wavenumber $k$, we call the instability a Turing-Hopf instability.
\end{definition}
Importantly, a Turing instability corresponds to linearly growing perturbation modes, while a Turing-Hopf instability implies growing perturbation modes that oscillate. For a system with $N=2$ species, it is well-known \cite{Murray2003-vr} that any instability must be a Turing instability and it occurs precisely when
\begin{equation}\label{eq:turing_2spec}
    J_{11}d_2+J_{22}d_1>2\sqrt{d_1d_2\mathrm{det}J}.
\end{equation}
For $N=3$ species, both Turing and Turing-Hopf instabilities can occur and linear stability has been analysed for some restricted choices of the matrix $J$ \cite{Manna2021-re, Korvasova2015-yx} or diffusivities $d_i$ \cite{Pearson1989-wl}. Moreover, partial conditions that specify the existence of diffusivities $d_i$ that give rise to instabilities for a fixed matrix $J$ have been provided \cite{Satnoianu2000-aj, Villar-Sepulveda2023-ny}.

%% Main result
\section{Main Result}\label{sec:Main}
In this paper, I find conditions that determine the presence of Turing and Turing-Hopf instabilities in a general reaction-diffusion system with $N=3$ species. To state the main result, consider the following two sets of inequalities:
\begin{equation}\label{eq:polynom_cond1}
\begin{aligned}
    0&<a_2^2-3a_1a_3,\\
    0&<a_2+\sqrt{a_2^2-3a_1a_3},\\
    0&<2a_2^3+2(a_2^2-3a_1a_3)^{3/2}-9a_1a_2a_3+27a_0a_3^2,\\
\end{aligned}
\end{equation}
and
\begin{equation}\label{eq:polynom_cond2}
\begin{aligned}
    b_1&<-\sqrt{4b_2b_0},\\
    3a_3(b_1+\sqrt{b_1^2-4b_2b_0})&\leq2b_0(a_2+\sqrt{a_2^2-3a_1a_3}),\\
    2b_0(a_2+\sqrt{a_2^2-3a_1a_3})&\leq3a_3(b_1-\sqrt{b_1^2-4b_2b_0}),\\
    g\left(\frac{-b_1-\sqrt{b_1^2-4b_2b_0}}{2b_2}\right)&\leq0,\\
    g\left(\frac{-b_1+\sqrt{b_1^2-4b_2b_0}}{2b_2}\right)&\leq0,\\
\end{aligned}
\end{equation}
where $g(y)=b_2y^2+b_1y+b_0$.
\begin{theorem}\label{thm:main}
    The spatially uniform steady-state $\mathbf{n}^*$ of \eqref{eq:pde_fix_diff} with $N=3$ species admits a Turing-Hopf instability if and only if the coefficients
    \begin{equation}
    \begin{aligned}\label{eq:coeff_a}
    a_3&=-(d_1+d_2)(d_2+d_3)(d_3+d_1)\\
    a_2&=(d_1+d_2)(d_1+d_2+2d_3)J_{33}+(d_2+d_3)(d_2+d_3+2d_1)J_{11}+(d_3+d_1)(d_3+d_1+2d_2)J_{22}\\
    a_1&=-[(d_1+d_2)(J_{11}J_{22}-J_{12}J_{21})+(d_2+d_3)(J_{22}J_{33}-J_{23}J_{32})+(d_3+d_1)(J_{33}J_{11}-J_{31}J_{13})]\\
    &-\mathrm{tr}J[d_1(J_{22}+J_{33})+d_2(J_{33}+J_{11})+d_3(J_{11}+J_{22})]\\
    a_0&=\mathrm{tr } J\mathrm{tr }[\mathrm{adj } J]-\mathrm{det }J
    \end{aligned}
    \end{equation}
    satisfy all the inequalities in \eqref{eq:polynom_cond1} and the coefficients
    \begin{equation}\label{eq:coeff_b}
    \begin{aligned}
        b_2&=d_1d_2+d_1d_3+d_2d_3\\
        b_1&=-[d_1(J_{22}+J_{33})+d_2(J_{33}+J_{11})+d_3(J_{11}+J_{22})]\\
        b_0&=\mathrm{tr }[\mathrm{adj } J]\\
    \end{aligned}
    \end{equation}
    do not satisfy at least one inequality in \eqref{eq:polynom_cond2}. Moreover, the steady-state $\mathbf{n}^*$ of \eqref{eq:pde_fix_diff} admits a Turing instability if and only if it does not admit a Turing-Hopf instability and the coefficients
    \begin{equation}\label{eq:coeff_tilde_a}
    \begin{aligned}
        \tilde{a}_3 &= -d_1d_2d_3\\
        \tilde{a}_2 &= d_1d_2J_{33}+d_2d_3J_{11}+d_3d_1J_{22}\\
        \tilde{a}_1 &= -[d_1(J_{22}J_{33}-J_{23}J_{32})+d_2(J_{33}J_{11}-J_{31}J_{13})+d_3(J_{11}J_{22}-J_{12}J_{21})]\\
        \tilde{a}_0 &= \mathrm{det }J
    \end{aligned}
    \end{equation}
    satisfy all the inequalities in \eqref{eq:polynom_cond1}.
\end{theorem}

\section{Auxiliary Results}
Since the proof of the main result relies on properties of the characteristic polynomial for $J(k^2)$, I start by proving auxiliary results about polynomials.
\begin{lemma}\label{lem:Polynomials2}
    The cubic function $p(\lambda)=\lambda^3-c_2\lambda^2+c_1\lambda-c_0$ has three complex roots. Moreover, assuming $c_2<0$, the following statements hold:
    \begin{enumerate}[(a)]
        \item At least one of the roots has a negative real part.
        \item All roots have negative real parts precisely when $c_1>0$, $c_0<0$ and $c_2c_1<c_0$.
        \item Two roots have negative real parts and one root is real positive precisely when $c_0>0$.
        \item If $c_1>0$ and $c_2c_1>c_0$, then one root is real negative and two roots are complex conjugate with positive real parts.
    \end{enumerate}
\end{lemma}
\begin{proof}
    By the fundamental theorem of algebra, the cubic function $p(\lambda)$ has three complex roots, say $\lambda_1$, $\lambda_2$ and $\lambda_3$. Moreover, 
    \begin{equation}
    \begin{aligned}
        c_2 &= \lambda_1+\lambda_2+\lambda_3\\
        c_1 &= \lambda_1\lambda_2+\lambda_2\lambda_3+\lambda_3\lambda_1\\
        c_0 &= \lambda_1\lambda_2\lambda_3.\\
    \end{aligned}
    \end{equation}
    To prove (a), notice that three roots with positive real parts have a positive sum of their real parts, that is $c_2>0$, contradictory to the assumption that $c_2<0$. Furthermore, (b) follows from the Routh-Hurwitz criterion.
    
    To prove (c), notice that if two roots $\lambda_1$ and $\lambda_2$ have negative real parts and one root $\lambda_3$ is real positive, then either $\lambda_1=\overline{\lambda_2} \notin \mathbb{R}$ or $\lambda_1,\lambda_2<0$. If $\lambda_1=\overline{\lambda_2} \notin \mathbb{R}$, then $c_0=|\lambda_1|^2\lambda_3>0$. If $\lambda_1,\lambda_2<0$, then $c_0=\lambda_1\lambda_2\lambda_3>0$. To prove the converse, notice that the three roots are either (i) all real or (ii) one is real and two are complex conjugates, say $\lambda_1$ and $\lambda_2$. In the case (i), $c_0=\lambda_1\lambda_2\lambda_3>0$ implies that an even number of roots must be negative. Therefore, there are two negative roots and one positive root, as otherwise, all roots positive contradictions $c_2 = \lambda_1+\lambda_2+\lambda_3<0$. In the case (ii), $c_0=|\lambda_1|^2\lambda_3>0$. Therefore, $\lambda_3>0$ and $\mathrm{Re} \lambda_1=\mathrm{Re} \lambda_2=(c_2-\lambda_3)/2<0$.

    To prove (d), notice that if $c_1>0$, $c_2<0$ and $c_2c_1>c_0$, then $c_0<0$. Moreover, the three roots are either (i) all real or (ii) one is real and two are complex conjugates, say $\lambda_1$ and $\lambda_2$. In the case (i), $c_0=\lambda_1\lambda_2\lambda_3<0$ and there must be an odd number of negative roots. If all three roots were negative, then (b) implies that $c_2c_1<c_0$, contradictory to $c_2c_1>c_0$. If two roots are positive and one is negative, then $p(\lambda)$ must attain a positive minimum
    \begin{equation}\label{eq:}
        \lambda_m=\frac{c_2+\sqrt{c_2^2-3c_1}}{3}>0,
    \end{equation}
    and $c_2^2>3c_1$. However, as $\sqrt{c_2^2-3c_1}<-c_2$ and $c_2<0$, $\lambda_m<0$, which implies that (i) leads to a contradiction. In the case (ii), $c_0=|\lambda_1|^2\lambda_3<0$ and $\lambda_3<0$. Moreover, if the complex conjugate pair of roots had negative real parts, then (b) implies that $c_2c_1<c_0$, contradictory to $c_2c_1>c_0$. Therefore, $\lambda_3<0$ and $\mathrm{Re}\lambda_1=\mathrm{Re}\lambda_2>0$.
\end{proof}
\begin{lemma}\label{lem:Polynomials}
    Let $g(y)=a_3y^3+a_2y^2+a_1y+a_0$ be a cubic function with $a_3<0$ and $a_0<0$, and let $h(x)=b_2y^2+b_1y+b_0$ be a quadratic function with $b_2>0$ and $b_0>0$. Then, there is a positive number $y'>0$ such that $g(y') > 0$ iff all inequalities in \eqref{eq:polynom_cond1} are satisfied. Furthermore, there is a positive number $y''>0$ such that $g(y'') > 0$ and $h(y'')>0$ iff all inequalities in \eqref{eq:polynom_cond1} are satisfied and at least one inequality in \eqref{eq:polynom_cond2} is not satisfied.
\end{lemma}
\begin{proof}
    Let $\mathrm{d} g/\mathrm{d} y = 3a_3y^2+2a_2y+a_1$. If $\mathrm{d} g/\mathrm{d} y=0$ has at most one real solution, then $\mathrm{d} g/\mathrm{d} y \leq 0$ and $g(0)=a_0 <0$ imply that $g(y)<0$ for all $y>0$, contradicting the existence of $y'$. If $\mathrm{d} g/\mathrm{d} y=0$ has two real solutions, it must be true that $0 < a_2^2-3a_3a_1$ (i) and the solutions are given by
    \begin{equation}
        y_\pm = \frac{-a_2 \pm \sqrt{a_2^2-3a_1a_3}}{3a_3}
    \end{equation}
    with $y_+ < y_-$. Since $a_3<0$, the solution $y_+$  (resp. $y_-$) corresponds to a local minimum (resp. maximum). If $y_-<0$, then $g(y)$ is monotone decreasing for $y \geq 0$ with $g(0)=a_0 <0$, contradicting the existence of $y'$ by the previous argument. Therefore, for $y'$ to exist, it must be true that $y_->0$, or equivalently $0<a_2+\sqrt{a_2^2-3a_1a_3}$ (ii). Finally, there is an $y'>0$ with $g(y')>0$ iff (i) and (ii) are satisfied and the value of $g(y_-)$ at the maximum $y_-$ is positive, or equivalently $0<2a_2^3+2(a_2^2-3a_1a_3)^{3/2}-9a_1a_2a_3+27a_0a_3^2$.

    Furthermore, there is $y''>0$ such that $g(y'')>0$ and $h(y'')>0$ iff all inequalities in \eqref{eq:polynom_cond1} are satisfied and the open interval $S_g=\{y|g(y)>0,y>0\}$ is not included in the closed interval $S_h=\{y|h(y) \leq 0,y>0$\}. Since $b_2>0$, the set $S_h$ either contains at most one point or $b_1 < -\sqrt{4b_2b_0}$ and $S_h=[y_1,y_2]$, where
    \begin{equation}
        y_{1,2}=\frac{-b_1\pm\sqrt{b_1^2-4b_2b_0}}{2b_2}.
    \end{equation}
    Then, $S_h$ contains $S_g$ precisely when $S_h=[y_1,y_2]$ for $y_1<y_2$, the maximum $y_-$ of $g(y)$ lies in $S_h$ and $g(y_{1,2}) \leq 0$. Expansion of these conditions gives rise to inequalities in \eqref{eq:polynom_cond2}.
\end{proof}

\section{Proof of the Main Result}
\begin{figure}
    \centering
    \includegraphics[width=0.7\textwidth]{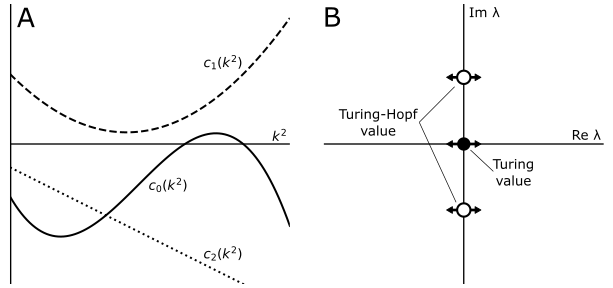}
    \caption{(A) Illustration of how coefficients of the characteristic equation vary with wavenumbers $k^2$ for a fixed choice of diffusivities $d_i$ and Jacobian matrix $J$, see \eqref{eq:coeff_char_eq}. (B) Definition of a Turing value and a Turing-Hopf value. Eigenvalues cross $\mathrm{Re}\lambda=0$ as $k^2$ is varied (indicated by arrows).}
   \label{fig:figure}
\end{figure}

To prove the main result, notice that the eigenvalues $\lambda$ of $J(k^2)$ in \eqref{eq:jacobian} are given by the characteristic equation
\begin{equation}
    p_{k^2}(\lambda)=\lambda^3-c_2(k^2) \lambda^2+c_1(k^2)\lambda-c_0(k^2)=0,
\end{equation}
where
\begin{equation}\label{eq:coeff_char_eq}
\begin{aligned}
c_2(k^2)&=\mathrm{tr } J(k^2) = -(d_1+d_2+d_3)k^2+\mathrm{tr } J\\
c_1(k^2)&=\mathrm{tr }[\mathrm{adj } J(k^2)] =b_2k^4+b_1k^2+b_0\\
c_0(k^2)&=\mathrm{det }J(k^2) =\tilde{a}_3k^6+\tilde{a}_2k^4+\tilde{a}_1k^2+\tilde{a}_0\\
c_2(k^2)&c_1(k^2)-c_0(k^2)=a_3k^6+a_2k^4+a_1k^2+a_0\\
\end{aligned}
\end{equation}
and the coefficients $a_j$, $b_j$ and $\tilde{a}_j$ are given by \eqref{eq:coeff_a}, \eqref{eq:coeff_b} and \eqref{eq:coeff_tilde_a}, see Fig. \ref{fig:figure}A. Moreover, the leading and constant coefficients of the polynomials in \eqref{eq:coeff_char_eq} have fixed signs: $-\sum_id_i<0$, $b_2>0$, $\tilde{a}_3<0$, $a_3<0$ (leading coefficients) and  $\mathrm{tr } J<0$, $b_0>0$, $\tilde{a}_0<0$ and $a_0<0$ (constant coefficients, follows by application of the Routh-Hurwitz criterion to the matrix $J(0)=J$). Consequently, $c_2(k^2)<0$ and Lemma \ref{lem:Polynomials2} applies to $p_{k^2}(\lambda)$ for any $k^2$. Moreover, the eigenvalues $\lambda$ vary continuously with $k^2$ and, by Lemma \ref{lem:Polynomials2}(b), have negative real parts at sufficiently small and sufficiently large values of $k^2$. Therefore, the steady-state $\mathbf{n}^*$ of \eqref{eq:pde_fix_diff} admits an instability precisely if some eigenvalue $\lambda$ crosses the boundary $\mathrm{Re}\lambda=0$ at some $k^2>0$, see Fig. \ref{fig:figure}B.

%If a real eigenvalue (resp. complex conjugate pair of eigenvalues) crosses the boundary $\mathrm{Re}\lambda=0$ at some $k^2>0$, then $k^2$ is called a Turing value (resp. Turing-Hopf value).
\begin{definition}
    If an eigenvalue $\lambda$ crosses the boundary $\mathrm{Re}\lambda=0$ at $\lambda=0$ (resp. pure imaginary value) for some $k^2>0$, then $k^2$ is called a Turing value (resp. Turing-Hopf value).
\end{definition}

For $N=3$ species, the Turing (resp. Turing-Hopf) values characterise the Turing (resp. Turing-Hopf) instabilities.
\begin{lemma}\label{lem:values_instab}
    The steady-state $n_i^*$ of \eqref{eq:pde_fix_diff} with $N=3$ species admits a Turing-Hopf instability if and only if there exists a Turing-Hopf value $\kappa^2$. Moreover, it admits a Turing instability if and only if there exists a Turing value $\kappa^2$ and no value $k^2$ is a Turing-Hopf value.
\end{lemma}
\begin{proof}
    Since instability occurs precisely when there exists a Turing or a Turing-Hopf value and the existence of a Turing-Hopf value implies a Turing-Hopf instability, it suffices to show that a Turing-Hopf instability implies the existence of a Turing-Hopf value. For a contradiction, suppose that there is a Turing-Hopf instability but not a Turing-Hopf value. Due to a Turing-Hopf instability, there is a complex conjugate pair of eigenvalues with positive real parts at some value $\kappa^2>0$. By Lemma \ref{lem:Polynomials2}(a), the remaining eigenvalue at $\kappa^2$ is negative real. Since eigenvalues vary continuously with $k^2$ and all eigenvalues have negative real parts for sufficiently small and sufficiently large $k^2$, the two eigenvalues that form a complex conjugate pair at $\kappa^2$ must cross the line $\mathrm{Re} \lambda=0$ as $k^2$ increases and decreases from $\kappa^2$. Since there is no Turing-Hopf value, there are either (i) at least three distinct Turing values or (ii) two distinct Turing values and the pair of eigenvalues collides at these Turing values. In the case of (i), the product of eigenvalues $c_0(k^2)$ has at least three distinct positive roots corresponding to the Turing values. However, $c_0(y)$ is a cubic with $c_0(0)=\tilde{a}_0<0$ and a negative leading coefficient, implying that it can admit at most two positive real roots. In the case of (ii), $c_0(y)$ has two positive real roots that coincide with the two Turing values. Moreover, as a pair of eigenvalues crosses $\lambda=0$ at the Turing values, the product of the eigenvalues $c_0(y)$ does not change a sign at the crossing. Consequently, the two Turing values are distinct positive roots and maxima of $c_0(y)$, contradicting that the cubic $c_0(y)$ can admit at most one positive maximum.
\end{proof}

Finally, as the signs of the leading and constant coefficients in \eqref{eq:coeff_char_eq} are fixed, Lemma \ref{lem:Polynomials} can be applied to the function $g(y)=c_0(y)$ or to the pair of functions $g(y)=c_2(y)c_1(y)-c_0(y)$ and $h(y)=c_1(y)$. In particular, the main result in Theorem \ref{thm:main} follows if I prove the following two results in Lemma \ref{lem:turing_hopf} and Lemma \ref{lem:turing}.
\begin{lemma}\label{lem:turing_hopf}
    The steady-state $\mathbf{n}^*$ of \eqref{eq:pde_fix_diff} with $N=3$ species admits a Turing-Hopf instability iff there is a value $y''>0$ such that $c_1(y'')>0$ and $c_2(y'')c_1(y'')-c_0(y'')>0$.
\end{lemma}
\begin{proof}
($\Rightarrow$) By Lemma \ref{lem:values_instab}, a Turing-Hopf instability implies the existence of a Turing-Hopf value $\kappa^2$ when a pair of complex conjugate eigenvalues $\lambda=\pm i\omega \in i \mathbb{R}\setminus\{0\}$ exists. In particular, $\omega^2c_2(\kappa^2)=c_0(\kappa^2)$ and $\omega^2=c_1(\kappa^2)$, implying that $c_2(\kappa^2)c_1(\kappa^2)=c_0(\kappa^2)$ and $0<c_1(\kappa^2)$.
By Lemma \ref{lem:Polynomials2}(a), the remaining eigenvalue must remain negative in the neighbourhood of wavenumbers around $\kappa$. As $k$ crosses $\kappa$, the distribution of eigenvalues changes from three eigenvalues with negative real parts to a single negative eigenvalue and a complex conjugate pair of eigenvalues with a positive real part. Therefore, by Lemma \ref{lem:Polynomials2}(b), $c_2(k^2)c_1(k^2)-c_0(k^2)$ changes from negative to positive and $c_1(\kappa^2)$ remains positive as $k$ crosses $\kappa$. Therefore, there must exist $y''>0$ such that $c_1(y'')>0$ and $c_2(y'')c_1(y'')-c_0(y'')>0$.

($\Leftarrow$) By Lemma \ref{lem:Polynomials2}(d), the existence of $y''>0$ such that $c_1(y'')>0$ and $c_2(y'')c_1(y'')-c_0(y'')>0$ implies that there is a complex conjugate pair of eigenvalues with a positive real part at $k=\sqrt{y''}$, corresponding to a Turing-Hopf instability.
\end{proof}

It is worth remarking that, for $N=3$ species, the existence of a Turing-Hopf instability not only ensures that there is a value $\kappa^2$ such that $J(\kappa^2)$ has a complex conjugate pair of eigenvalues with a positive real part but also that these eigenvalues have the largest real part (Lemma \ref{lem:Polynomials2}(a)), implying that the oscillatory growth of the perturbation mode with wavelength $\kappa$ provides the dynamically dominant behaviour.

\begin{lemma}\label{lem:turing}
    The steady-state $\mathbf{n}^*$ of \eqref{eq:pde_fix_diff} with $N=3$ species admits a Turing instability iff it does not admit a Turing-Hopf instability and there is a value $y'>0$ such that $c_0(y')>0$.
\end{lemma}
\begin{proof}
    ($\Rightarrow$) By Lemma \ref{lem:values_instab}, a Turing instability implies the absence of a Turing-Hopf instability and the existence of a Turing value $\kappa^2$. At $\kappa^2$, one or two eigenvalues $\lambda$ pass through $\lambda=0$, while at least one eigenvalue has a negative real part by  Lemma \ref{lem:Polynomials2}(a). If two eigenvalues pass through $\lambda=0$ simultaneously, then the product of eigenvalues $c_0(k^2)$ remains unchanged at $\kappa^2$, implying that $\kappa^2$ is a local maximum of $c_0(k^2)$ with $c_0(\kappa^2)=0$. Since $c_0(k^2)$ is a cubic, $\kappa^2$ is a global maximum and $c_0(k^2)<0$ for any $k^2 \neq \kappa^2$. Consequently, there is no other Turing value. Since the two eigenvalues that cross into the region $\mathrm{Re} \lambda>0$ must leave this region at sufficiently large and sufficiently small $k^2$, there must exist a Turing-Hopf value, contradicting the existence of a Turing instability due to Lemma \ref{lem:values_instab}. Therefore, only one eigenvalue can pass through $\lambda=0$. When this happens, the remaining two eigenvalues either have negative real parts or one is real positive and the other real negative. Consequently, there is a value $y'=k^2$ in the neighbourhood of $\kappa^2$ such that there is one real positive eigenvalue and two eigenvalues with negative real parts. By Lemma \ref{lem:Polynomials2}(c), it follows that $c_0(y')>0$. 

    ($\Leftarrow$) If there is a value $\kappa^2>0$ with $c_0(\kappa^2)>0$, then Lemma \ref{lem:Polynomials2}(c) implies that there is one real positive eigenvalue and two eigenvalues with negative real parts at $\kappa^2>0$. As $k^2$ decreases from $\kappa^2$, some eigenvalue must pass through $\textrm{Re } \lambda = 0$ since all eigenvalues have negative real parts for sufficiently small $k^2$. Moreover, at least one of the eigenvalues that pass through $\textrm{Re } \lambda = 0$ must be real since the parity of real eigenvalues that pass through $\textrm{Re } \lambda = 0$ must coincide with the parity of eigenvalues in the region $\textrm{Re } \lambda>0$ and there is one such eigenvalue at $\kappa^2$. Therefore, there exists a Turing value. By Lemma \ref{lem:values_instab}, the existence of a Turing value and the absence of a Turing-Hopf instability implies the existence of a Turing instability.
\end{proof}

\section{Conclusions}
I derived the conditions for Turing patterns to form in a general reaction-diffusion system with three interacting species (Theorem \ref{thm:main}). The Turing patterns form due to the instability of the homogeneous stationary state to spatial perturbations. I showed that the pertubations can either grow steadily (Turing instability) or with oscillations in time (Turing-Hopf instability) and analyzed when each instability type occurs (Theorem \ref{thm:main}).

\section{Acknowledgements}
I thank Philip Maini and Vaclav Klika for their useful comments on this manuscript. V.P. was supported by the Mathematical Institute Scholarship.

\section{Competing Interests}
The author declares no competing interests.

 \bibliographystyle{elsarticle-num} 
 \bibliography{cas-refs}





% \end{thebibliography}
\end{document}